\documentstyle[aps,epsf,twocolumn]{revtex}
\begin{document}
\title{On entanglement capabilities in infinite dimensions\\
or \\
multi-dimensional entangled coherent states}
\author{S.J. van Enk\\
Bell Labs, Lucent Technologies\\
600-700 Mountain Ave\\
Murray Hill NJ 07974}
\maketitle
\begin{abstract}
An example is given of an interaction that produces an infinite amount of entanglement in an infinitely short time, but only a finite amount in longer times. 
The interaction arises from a standard Kerr nonlinearity and a 50/50 beamsplitter, and the initial state is a coherent state.
For certain finite interaction times multi-dimensional generalizations of entangled coherent states are generated, for which we construct a teleportation protocol. Similarities between probabilistic teleportation and unambiguous state discrimination are pointed out. \end{abstract}
\medskip

Entangling operations are necessary for universal quantum computation \cite{nielsen} but also play an important role in quantum communication. 
For example, teleportation \cite{telep}, entanglement distillation \cite{distill} and quantum repeaters \cite{repeat} all rely on entangling operations.  

Questions concerning the entangling power of given unitary operations or of given Hamiltonians are thus relevant from both theoretical and practical points of view and have been considered recently in the context of finite-dimensional systems \cite{cirac,zanardi,leung}. 
But for the electromagnetic field, which is clearly the system of choice for quantum communication, the  associated Hilbert space is infinite. 
Entanglement in infinite Hilbert spaces has peculiar properties \cite{clifton,eisert,werner}. For example, there is always a state with an arbitrarily large amount of entanglement arbitrarily close to a separable state. 
Some of these anomolies can be mitigated by imposing energy constraints on the states considered \cite{eisert}. Here we confirm this behavior of entanglement in infinite-dimensional systems. In fact, we show that a standard nonlinear optics interaction, arising from a Kerr nonlinearity, followed by a simple interaction with a beamsplitter is capable of generating an arbitrarily large amount of entanglement $E=\log_2 M$ in an arbitrarily short time $\tau={\cal O}(1/M)$, starting from a coherent state with energy $|\alpha|^2={\cal O} (M^2)$. Note that squeezing or downconversion, in contrast, are only capable of generating Gaussian states from coherent states, whose entanglement after passing beamsplitters (and other linear-optics elements) is much better behaved \cite{beam}. For instance, the entanglement of a two-mode squeezed state simply increases monotonically with the squeezing parameter, which in turn increases with the interaction time.

It turns out that for certain specific finite interaction times
the interaction we will consider generates multi-dimensional generalizations of so-called entangled coherent states.
Entangled coherent states are of the (unnormalized) form
\begin{equation}\label{phi}
|\Phi_2\rangle=|\alpha\rangle|\alpha\rangle +\exp(i\phi)|-\alpha\rangle|-\alpha\rangle,
\end{equation}
with $|\alpha\rangle$ a coherent state with amplitude $\alpha$.
As far as the author knows, this type of states was discussed first in 1986 \cite{first}, and the name entangled coherent states was coined in Ref.~\cite{sanders}. 
There has been a lot of interest in the quantum-information processing capabilities of such states after it was found that the states with $\phi=\pi$ possess exactly one ebit of entanglement\cite{hirota} irrespective of the amplitude $\alpha$. Teleportation\cite{enk}, entanglement purification\cite{puri}, Bell-inequality violations\cite{Bell}, and universal quantum computing\cite{comp} have all been discussed in this context. 

Here we study a particular multi-dimensional generalization of the states (\ref{phi}) of the (unnormalized) form,
\begin{equation}\label{M}
|\Phi_M\rangle=\frac{1}{\sqrt{M}}\sum_{q=0}^{M-1} 
\exp(i\phi_q)
|\alpha\exp\big(\frac{2\pi qi}{M}\big)\rangle
|\alpha\exp\big(\frac{2\pi qi}{M}\big)\rangle,
\end{equation}
with $M>1$ an integer. These states should not be confused with multi-{\em mode} generalizations of entangled coherent states of the form
\begin{equation}\label{Psi}
|\Psi_M\rangle=(|\alpha\rangle)^{\otimes M}
+\exp(i\phi)(|-\alpha\rangle)^{\otimes M},
\end{equation}
which can be trivially generated from a state (\ref{phi}) by mixing it with the vacuum on beamsplitters, and do not possess more than one ebit of bipartite entanglement. The states (\ref{M}), on the other hand, are still 2-mode states, but may contain more than one ebit of entanglement. We choose to restrict the form of (\ref{M}) to containing only {\em symmetric} coherent states with coefficients of {\em equal} magnitude, because that is the type of states that can be generated by propagating a coherent state through a medium with a Kerr nonlinearity (see below). Such states  potentially have $\log_2 M$ ebits of bipartite entanglement. In particular, they reach that limit for large $\alpha/M$.

Consider the following unitary operator
\begin{equation}\label{U}
U_{A,B}(\tau)=\exp\big(\frac{\pi}{4} (a^{\dagger}b-b^{\dagger}a)\big)\exp(-i\tau a^{\dagger 2}a^2),
\end{equation}
where $\tau$ represents a dimensionless time, and $a^\dagger,a,b^\dagger,b$ are the  creation and annihilation operators for two modes $A$ and $B$, respectively. Physically, the first term corresponds to a 50/50 beamsplitter, the second describes the propagation of mode $A$  through a Kerr medium, for which the effective Hamiltonian is
\begin{equation}
H=\hbar \chi a^{\dagger 2}a^2,
\end{equation}
with $\chi$ a rate determined by the appropriate third-order nonlinear susceptibility of the medium. (Note that such an interaction has been considered for the generation of entangled coherent states\cite{generate}). We will be interested in the entangling capabilities of $U_{A,B}(\tau)$. The class of initial states of modes $A$ and $B$ is chosen from product states such that for $\tau=0$ no entanglement is generated. It is easy to see that any product of two coherent states will fit the bill. Here we take a subclass, namely $|\Psi\rangle_{A,B}=|\beta\rangle_A|0\rangle_B$ with arbitrary $\beta$, as initial states.
It is straightforward to expand the state
\begin{equation}\label{S}
|\Psi(\tau)\rangle_{A,B}=U_{A,B}(\tau)|\beta\rangle_A|0\rangle_B
\end{equation}
in number states and subsequently evaluate the entanglement $E_{A,B}(\tau)$
between modes $A$ and $B$ as a function of time. Here, however, we give a more elegant description valid at certain times $\tau$. This treatment is based on Ref.~\cite{tara}.
First, consider the second term in Eq.~(\ref{U}), and write it as
\begin{equation}
U_A(\tau)=\exp(-i \tau \hat{N}_A(\hat{N}_A-1)),
\end{equation}
where $\hat{N}_A=a^{\dagger} a$.
The operator $U_A$ becomes periodic in $N$ with period $M$ (that is, it becomes invariant under $\hat{N}_A\rightarrow \hat{N}_A+M$) at times $\tau=\pi/M$ if $M$ is an odd integer.
This implies one can write down Fourier series as follows\cite{tara}
\begin{equation}
\exp\big( \frac{-i\pi}{M}\hat{N}(\hat{N}-1)  \big)=\sum_{q=0}^{M-1} f_q^{(o)}\exp
\big( \frac{-2i\pi q}{M}\hat{N}\big).
\end{equation}
Similarly, for even values of $M$ one has
\begin{equation}\label{even}
\exp\big( \frac{-i\pi}{M}(\hat{N}+M)^2  \big)=
\exp\big( \frac{-i\pi}{M}\hat{N}^2  \big),
\end{equation}
so that we can expand
\begin{equation}
\exp\big( \frac{-i\pi}{M}\hat{N}^2  \big)=\sum_{q=0}^{M-1} f_q^{(e)}\exp
\big( \frac{-2i\pi q}{M}\hat{N}\big).
\end{equation}
The coefficients $f_q$ are not explicitly evaluated in Ref.~\cite{tara}, but one can 
actually derive them (see below \cite{proof}),
\begin{eqnarray}\label{MM}
f_q^{(o)}&=&\frac{1}{\sqrt{M}}\exp\big(\frac{\pi iq(q+1)}{M}  \big)
\exp\big(\frac{-\pi iK(K+1)}{M}  \big),\nonumber\\
f_q^{(e)}&=&\frac{1}{\sqrt{M}}\exp\big(\frac{\pi iq^2}{M} \big) \exp(-\pi i/4),
\end{eqnarray}
where in the first line $K$ is such that $M=2K+1$ for odd $M$.
If one starts with a coherent state of mode $A$ at time $\tau=0$, $|\psi(0)\rangle=|\beta\rangle$,
this then immediately leads to the following time evolution under $U_A$:
\begin{eqnarray}
U_A(\pi/M)|\beta\rangle_A
&=&\sum_{q=0}^{M-1}f_q^{(o)}|\beta\exp(-2\pi iq/M)\rangle\nonumber\\
{\rm for}\,\,M\,\,{\rm odd},\nonumber\\
U_A(\pi/M)|\beta\rangle_A
&=&\sum_{q=0}^{M-1}f_q^{(e)}|\beta\exp(\pi i(1-2q)/M)\rangle\nonumber\\
{\rm for}
\,\,M\,\,{\rm even}\end{eqnarray}
If one subsequently takes these states and splits them on a 50/50 beamsplitter with the vacuum, the output state is an entangled state of the form (\ref{M}),
\begin{equation}\label{Mo}
|\Phi_M\rangle=\sum_{q=0}^{M-1}f_q^{(o)} |\alpha\exp(-2\pi iq/M)\rangle
 |\alpha\exp(-2\pi iq/M)\rangle,
\end{equation}
for $M$ odd 
with $\alpha=\beta/\sqrt{2}$, and
\begin{equation}\label{Me}
|\Phi_M\rangle=\sum_{q=0}^{M-1}f_q^{(e)} |\alpha\exp(-2\pi iq/M)\rangle
|\alpha\exp(-2\pi iq/M)\rangle,
\end{equation}
for $M$ even, where now $\alpha=\beta\exp(i\pi/M)/\sqrt{2}$.
These are the states $|\Psi(\pi/M)\rangle_{A,B}$ of Eq.~(\ref{S})
we were looking for.

Now consider the entanglement between modes $A$ and $B$ in the states (\ref{Mo}) and (\ref{Me}).
In the limit of large $\alpha$, more precisely, for $\alpha/M\gg 1$, the coherent states appearing in these superpositions become orthogonal. The entangled states, therefore, are already written in their Schmidt decomposition form, and it is straightforward to calculate their entanglement. In fact, since all coefficients $f_q$ have the same magnitude $1/\sqrt{M}$, one sees one ends up with a maximally entangled state in $M$ dimensions (spanned by the $M$ symmetric coherent states), with $E_{A,B}(\pi/M)=\log_2 M$ ebits.
This fact is paradoxical at first sight, since the entanglement increases with $M$, that is, with {\em decreasing} interaction time. However, this paradox is resolved easily by noting that for fixed $\alpha$ the coherent states $|\alpha \exp(\pi i q/M)\rangle$ become nonorthogonal for sufficiently large $M$ so that the entanglement is in fact smaller than $\log_2(M)$. Thus, just as in \cite{eisert}, an energy constraint saves us from the more embarrassing peculiarities of entanglement in infinite dimensions.
In fact, there is an optimal time $\tau$ for which the entanglement is maximal for fixed $\alpha$.
This is illustrated in Figure~1. Here, for several finite values of $\alpha$ we evaluate the entanglement numerically by expanding the reduced density matrix of either of the two subsystems in the number-state basis.

For $\alpha$ small, there is no entanglement as the state $|\Phi_M\rangle$ reduces to the vacuum state. This follows immediately from the fact that the evolution operator $U_{A,B}$ commutes with the sum of the number operators, $a^\dagger a+b^\dagger b$, so that the total photon number distribution does not change.
\begin{figure}\label{f1} \leavevmode
\epsfxsize=8cm \epsfbox{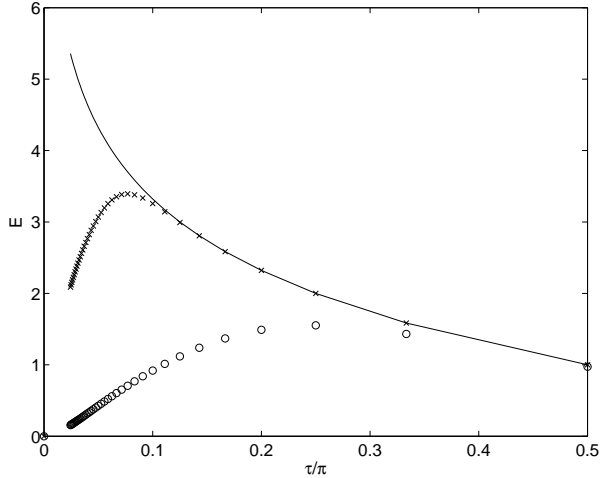} \caption{
Entanglement as a function of time $\tau/\pi$ for $|\alpha|^2=1$ (circles) and $|\alpha|^2=10$ (crosses), where $\tau/\pi=1/M$ with $M$ an integer $>1$. At these times the entangled state is an $M$-dimensional entangled coherent state.
The function $f(\tau)=\log_2(\pi/\tau)$ is given as reference.
}
\end{figure}

For a large subset of pairs of values of $\alpha$ and $M$, 
the entangled states (\ref{Mo}) or (\ref{Me}) possess more than one ebit of entanglement. As such, they can be used for teleporting at least one qubit.
Here we give a simple protocol for even values of $M$ that works perfectly in the limit of large $\alpha$, while for smaller $\alpha$ it works only partially, namely,  with a probability less than unity and with a fidelity less than unity. It generalizes the protocol of \cite{enk}.

Suppose Alice and Bob share an entangled state $|\Phi_M\rangle_{A,B}$
of the form (\ref{Me}) ($M$ is even)
between modes $A$ and $B$, and Alice possesses an
arbitrary state of the form
\begin{equation}
|\psi\rangle_C=\sum_{q=0}^{M-1} Q_q |\alpha \exp(-2\pi iq/M)\rangle_C,
\end{equation}
that she wishes to teleport to Bob. 
Alice first uses beamsplitters to make $L=M/2$ ``diluted'' copies of both the state to be teleported (ending up in modes $C_k$ for $k=0\ldots L-1$) and of her half of the entangled state (ending up in modes $A_k$ for $k=0\ldots L-1$) by the process
\begin{eqnarray}
|\alpha\exp(i\phi)\rangle(|0\rangle )^{\otimes L-1}\mapsto 
(|\alpha\exp(i\phi)/\sqrt{L}\rangle)^{\otimes L}.
\end{eqnarray}
Then she applies a phase shift over an angle $\phi_k=2k\pi/M$ to the modes $A_k$ and, in order to perform her Bell measurement,  subsequently combines the modes $C_k$ and $A_k$ on $L$ 50/50 beamsplitters. If we call the output modes $G_k$ and $H_k$ for $k=0\ldots L-1$, 
the resulting state is 
\begin{eqnarray}\label{fin}
&&\sum_{q=0}^{M-1}\sum_{p=0}^{M-1}\otimes_{k=0}^{L-1}
f_q^{(e)} Q_p |\alpha\exp(-2\pi iq/M)\rangle_B
\nonumber\\
&\otimes& |\alpha [\exp(-2\pi iq/M)-\exp(-2\pi i(p+k)/M)]/\sqrt{2L}\rangle_{G_k}
\nonumber\\
&\otimes&
|\alpha [\exp(-2\pi iq/M)+\exp(-2\pi i(p+k)/M)]/\sqrt{2L}\rangle_{H_k}.
\end{eqnarray}
Alice now performs photon-number measurements on all $2L=M$ output modes.
She cannot find a nonzero number in every mode. But suppose she finds nonzero numbers of photons in all but one mode, say, mode $H_m$. Then the only terms that survive the sums over $q$ and $p$ in 
(\ref{fin}) are those for which $\exp(-2\pi iq/M)+\exp(-2\pi i(p+m)/M)=0$, that is,
$p+m=q+L$ modulo $M$. The state at Bob's side reduces to (unnormalized)
\begin{equation}\label{Bob}
\sum_{q=0}^{M-1}f_{q}^{(e)}Q_{q\oplus L-m}\exp\big(\frac{-2\pi iq N_{{\rm tot}}}{M}\big) |\alpha\exp(-2\pi iq/M)\rangle_B,
\end{equation} 
with $\oplus$ denoting addition modulo $M$, and where $N_{{\rm tot}}$ is the total number of photons detected by Alice. Alice communicates to Bob which mode contained no photons, and Bob then applies the appropriate unitary transformation. Here, with $H_m$ being the empty mode, he applies
\begin{equation}
U_B=\exp\big(\frac{-2\pi i \hat{N}_B(L-m)}{M} \big)
\end{equation}
to his state to obtain (unnormalized)
\begin{eqnarray}\label{Bob2}
&&\sum_{q=0}^{M-1}\exp\big(\frac{\pi i(q-L\oplus m)^2}{M} \big)  
\exp\big(\frac{-2\pi iq N_{{\rm tot}}}{M}\big) 
\nonumber\\
&&Q_q |\alpha\exp(-2\pi iq/M)\rangle_B.
\end{eqnarray} 
This ensures that the coefficients $Q_q$ are in front of the corresponding states $|\alpha\exp(-2\pi iq/M)\rangle_B$. But what remains to complete teleportation would be a phase shift operation
\begin{eqnarray}
&&|\alpha\exp(-2\pi iq/M)\rangle\mapsto\nonumber\\
&&\exp\big(\frac{-\pi i(q^2+2q(m-L-N_{{\rm tot}}))}{M} \big)  
|\alpha\exp(-2\pi iq/M)\rangle.
\end{eqnarray}
This operation is in general unitary only in the limit of large $\alpha/M$.
For finite $\alpha/M$ one is thus able to perform teleportation only approximately.
Moreover, in that case the probability to find nonzero numbers of photons in every mode but one will be less than unity. 

The measurement performed by Alice is in fact the same as that needed for unambiguous state discrimination (USD) measurements on symmetric coherent states \cite{usd}. The only difference is that the USD measurement would have to be performed with a coherent state of {\em known} phase, not with half of the entangled state, in which the phase is basically unknown, that is, a mixture of $M$ values. This difference arises because for teleportation it is crucial that Alice's measurement does not reveal any information about the identity of the state to be teleported. In both USD and probabilistic teleportation the success probability may be less than unity (it becomes unity only in the limit of large $\alpha/M$), but one does know when it failed.

In conclusion, we considered the entangling capabilities of the unitary operator $U_{A,B}(\tau)$ of Eq.~(\ref{U}), which arises from a standard Kerr nonlinearity and simple linear optics.
Starting from coherent states, $M$-dimensional symmetric generalizations of entangled coherent states are generated for arbitrary integers $M$.
We constructed a teleportation protocol with these states that uses only linear optics and teleports states chosen from an appropriate $M$-dimensional Hilbert space.

Moreover, we found that an arbitrarily large amount  of entanglement $\log_2 M$ ebits can be created in an arbitrarily short time $\tau=\pi/M$. This is surprising as in finite dimensions not only the entanglement is finite (obviously) but also the rate of production of entanglement \cite{leung}. Thus one cannot create any finite amount of entanglement in an arbitrarily short time in a finite-dimensional space.

I thank Debbie Leung for pointing out this surprising contrast between the present result and that of Ref.~\cite{leung}.


\begin{thebibliography}{99}
\bibitem{proof}Here we prove the result (\ref{MM}) for $M$ even.
The case of odd $M$ is left to the reader as an exercise.

Apply the inverse Fourier inverse to Eq.~(\ref{even}) to get
\begin{equation}\label{proof1}
f_q^{(e)}=\frac{1}{M}\sum_{k=0}^{M-1}
\exp\big(\frac{2\pi iqk}{M}\big)\exp\big(\frac{-i\pi k^2}{M}\big).
\end{equation}
For $q=0$ this sum can be evaluated using relations (16.1.1) and (19.1.1) from 
Ref.\cite{tables}:
\[\sum_{k=0}^{K-1}\cos\big(\frac{2\pi k^2}{K}\big)=\frac{1}{2}\sqrt{K}(1+\cos(\pi K/2))+\sin(\pi K/2)),\]
and
\[\sum_{k=0}^{K-1}\sin\big(\frac{2\pi k^2}{K}\big)=\frac{1}{2}\sqrt{K}(1+\cos(\pi K/2))-\sin(\pi K/2)).\]
Namely, one easily finds by substituting $K=2M$
\begin{equation}
f_0^{(e)}=\frac{1-i}{\sqrt{2M}}.
\end{equation}
Subsequently one notices that the summand of Eq.~(\ref{proof1})
can be rewritten as
\[
\exp\big(\frac{2\pi iqk}{M}\big)\exp\big(\frac{-i\pi k^2}{M}\big)
=
\exp\big(\frac{i\pi q^2}{M}\big)
\exp\big(\frac{-i\pi (k-q)^2}{M}\big).
\]
This immediately leads to $f_q^{(e)}=f_q^{(0)}\exp\big(\frac{i\pi q^2}{M}\big)$, and Eq.~(\ref{MM}) for $M$ even follows in turn.
\end{thebibliography}
\end{document}